\documentclass[12pt]{article}
\usepackage{amssymb}
\usepackage{amsmath}
\usepackage{latexsym}
\usepackage{longtable}
\usepackage{epsfig}
\usepackage{graphicx,bbm,psfrag}
\graphicspath{{images/}}
\begin{document}
\newtheorem{Lemma}{Lemma}
\newtheorem{Proposition}[Lemma]{Proposition}
\newtheorem{Corollary}[Lemma]{Corollary}
\newtheorem{Theorem}[Lemma]{Theorem}
\newenvironment{Proof}{{\bf Proof:}}{\hspace*{\fill}$\Box$}
%\begin{titlepage}
\begin{center}
{\Large\bf On the Integrated Form of the BBGKY\medskip\\
Hierarchy for Hard Spheres\bigskip}\\
{\large Herbert Spohn}\bigskip\\
Theoretische Physik, Ludwig-Maximilians-Universit\"{a}t,\\
Theresienstra{\ss}e 37, 80333 M\"{u}nchen, Germany\bigskip\\
March 1985\bigskip\\
\end{center}
{\it A note on the history} (May 2006): In my book ``Large Scale Dynamics
of Interacting Particles'' [S] I refer to an unpublished note
from early 1985 on the BBGKY hierarchy for hard spheres. My main
point there was to provide a direct probabilistic proof for the
time-integrated version of the hierarchy. Over recent years there
has been repeated interest in this derivation, which encourages me
to make my note public. I decided to leave it in its original form
including likely inaccuracies. The
work of R. Illner and M. Pulvirenti [IP] appeared in September 1985, 
see also the book by
C. Cercignani, R. Illner, and M. Pulvirenti [CIP], who prove 
the same result using special flow representation and methods from 
the theory of differential operators.\smallskip\\
{[S]}  H. Spohn, Large Scale Dynamics of Interacting
Particles, Texts and Monographs in Physics, Springer-Verlag,
Heidelberg, 1991.\smallskip\\
{[IP]} R. Illner and M. Pulvirenti, A derivation of the
BBGKY-hierarchy for hard sphere particle systems,  Transport Theory and
Stat. Phys. \textbf{16}, 997--1012 (1987), preprint DM-388-IR,
September 1985.\smallskip\\
{[CIP]} C. Cercignani, R. Illner, and M. Pulvirenti, The
Mathematical Theory of Dilute Gases, Applied Mathematical Sciences
\textbf{106}, Springer-Verlag, New York, 1994.\smallskip\\
\rule{150mm}{1pt}\bigskip\\
The aim of these notes is to establish the integrated form of the
BBGKY hierarchy for hard spheres as used by O.E. Lanford \cite {1}
in his proof of the validity of the Boltzmann equation in the
Boltzmann--Grad limit. The idea of a direct probabilistic proof is
inspired by a paper of R. Lang and X.X. Nguyen \cite{2}.

We denote by $x_j = (q_j,p_j) \in \Lambda \times {\mathbb R}^3$
position and momentum of the $j$--th particle. The hard spheres have
diameter $a$ (and mass one). They are confined to the region
$\Lambda$. $\Lambda$ is bounded and has a ``smooth'' boundary
$\partial \Lambda$. Conditions on $\partial \Lambda$ ensuring the
existence of the hard sphere dynamics are given in the thesis of K.
Alexander \cite{3}, p. 13/14, and we assume the validity of these
conditions here. We have exactly $N$ particles, $j=1,\ldots,N.$  The
$n$-particle phase space, $n=1,2,\ldots, N,$ is
\setcounter{equation}{0}
\begin{eqnarray}\label{1}
&&\hspace{-10pt} \Gamma_n = \big\{ (x_1,\ldots, x_n) \in (\Lambda
\times {\mathbb
R}^3)^n \big| ~|q_i-q| \ge a/2 ~\mbox{for all}~\nonumber\\
 &&\hspace{20pt}q\in \partial \Lambda,
|q_i - q_j| \ge a ~, i,j=1,\ldots, n, i\not= j\big\}\,.
\end{eqnarray}

In a  collision of two hard spheres incoming and outgoing momenta
transform into each other as
\begin{eqnarray}\label{2}
&&\hspace{-28pt}
p_i^\prime = p_i - \hat\omega[\hat\omega \cdot (p_i-p_j)]\,,\nonumber\\
&&\hspace{-28pt} p_j' = p_j + \hat\omega[\hat\omega \cdot
(p_i-p_j)]\,,
\end{eqnarray}
$i\not=j,~\mbox{with}~ \hat\omega\in S^2$. At the wall particles are
specularly reflected,
\begin{equation}\label{3}
p_j^\prime = p_j - 2 \hat n(q_j) [\hat n (q_j)\cdot p_j]\,,
\end{equation}
where $\hat n(q_j)$ is the unit outward normal at the point of
contact. For the construction of the hard sphere dynamics we
refer to Alexander$^\prime$s thesis.

We remove once and for all from $\Gamma_n$ the set of points which
in the course of time run into either a grazing or a multiple
collision. The phase space with these points removed is denoted by
$\Gamma_n^\ast$. $\Gamma_n \setminus\Gamma_n^\ast$ has Lebesgue
measure zero. Then, for all $t\in \mathbb R$ and for every point
$(x_1,\ldots, x_n) \in \Gamma_n^\ast$, the flow
\begin{equation}\label{4}
t \mapsto T_t^{(n)}(x_1,\ldots, x_n)\in \Gamma_n^\ast
\end{equation}
is well defined. In particular $\Gamma_n^\ast$ is invariant under
$T_t^{(n)}$.

If incoming and outgoing momenta are identified (also at collisions
with the wall), then $T_t^{(n)}$ is continuous in $t$, i.e. for all
$(x_1,\ldots, x_n)\in \Gamma_n^\ast$ one has
\begin{equation}\label{5}
\lim\limits_{t\to 0} ~ T_t^{(n)}(x_1,\ldots, x_n) = (x_1,\ldots,
x_n)\,.
\end{equation}
Here we will {\it not} identify incoming and outgoing momenta, i.e.
we regard them as distinct phase points. For $ (x_1,\ldots, x_n)\in
\Gamma_n^\ast$ the map $ t \mapsto T_t^{(n)}(x_1,\ldots, x_n)$ is
then piecewise continuous and we have to distinguish between the
limit from the future $(+)$ and from the past $(-)$ defined by
\begin{equation}\label{6}
 T_{t\pm}^{(n)}(x_1,\ldots, x_n) = \lim_{\varepsilon\to 0,\varepsilon >
 0}\,
T_{t\pm\varepsilon}^{(n)}(x_1,\ldots, x_n)\,.
\end{equation}
If the added signs $\pm$ are omitted, it is understood that the
quantity in question is independent of the way the limit is taken.

A function $\rho_n: \Gamma_n \to \mathbb R$ is continuous along
trajectories of $T_t^{(n)}$ on $\Gamma_n^\ast$, if for all $
(x_1,\ldots, x_n)\in \Gamma_n^\ast$
\begin{equation}\label{7}
\lim_{t\to 0} ~\rho_n (T_t^{(n)}(x_1,\ldots, x_n)) =
\rho_n(x_1,\ldots,x_n)\,,
\end{equation}
where both the limit from the future and the past are understood.
This implies then that for all $(x_1,\ldots, q_i, p_i, \ldots, q_i+a
\hat \omega, p_j, \ldots, x_n) \in \Gamma_n^\ast$ one has
\begin{eqnarray}\label{8}
&&\hspace{0pt}\rho_n (x_1,\ldots, q_i, p_i, \ldots, q_i+a \hat
\omega, p_j, \ldots,
x_n)\nonumber\\
&&\hspace{0pt}= \rho_n (x_1,\ldots, q_i, p_i^\prime, \ldots, q_i+a
\hat \omega, p_j^\prime, \ldots, x_n)
\end{eqnarray}
and similarly for collisions with the wall.

For future convenience we define some sets: Let
\begin{eqnarray}\label{9}
&&\hspace{-20pt}\Gamma_{N-n}  (x_1,\ldots, x_n)= \big\{
(x_{n+1},\ldots, x_N)\in
\Gamma_{N-n} \big| ~|q_i-q_j| \ge a,\nonumber\\
&&\hspace{90pt}\mbox{for}~ i=1, \ldots, n ~ \mbox{and}~ j=n+1,
\ldots, N\big\}
\end{eqnarray}
for $(x_1, \ldots, x_n) \in \Gamma_n^\ast $ and
\begin{equation}
\Gamma_{N-n}(x_1,\ldots,x_n) = \emptyset\label{10}
\end{equation}
otherwise. Let
\begin{equation}\label{11}
\Omega_j (x_1,\ldots,x_n, p_{n+1}) = \big\{\hat\omega \in S^2 \big|
(x_1,\ldots, x_n, q_j + a\hat\omega, p_{n+1}) \in
\Gamma_{n+1}^\ast\big\} \subset S^2
\end{equation}
for $j=1,\ldots,n$, $(x_1,\ldots,x_n) \in \Gamma_n^\ast$, and 
$p_{n+1} \in {\mathbb R}^3$ and let
\begin{equation}\label{12}
\Omega_j(x_1,\ldots, x_n, p_{n+1}) = \emptyset
\end{equation}
otherwise. We define two subsets, $\Omega_{j\pm}$, of $\Omega_j$ by
\begin{equation}\label{13}
\Omega_{j\pm} (x_1,\ldots,x_n, p_{n+1})= \big\{\hat\omega \in
\Omega_j(x_1,\ldots,x_n, p_{n+1})\big|\hat \omega \cdot  (p_{n+1}-
p_j) > 0~ (<0)ß\big\}\,.
\end{equation}

After these preparations we can state our assumptions on the
initial $(t=0)$ measure.\medskip\\
Let $P$ be the initial probability measure on $\Gamma_N$. $P$ is
assumed to satisfy:\\
(i) \quad$P$ is symmetric in the particle labels.\\
(ii) \quad $P$ has a density,
\begin{equation}\label{14}
P(dx_1 \ldots dx_N)=f_N(x_1,\ldots, x_N) dx_1 \ldots dx_N .
\end{equation}
(iii) \quad $f_N$ is bounded by the canonical equilibrium
distribution, i.e. there exist constants $c,\beta > 0$ such that
\begin{equation}\label{14a}
f_N(x_1,\ldots, x_N) \le c \prod_{j=1}^N h_\beta (p_j)\,
\end{equation}
on $\Gamma_N$, where $h_\beta(p) = (\frac{\beta}{2 \pi})^{3/2}
e^{-\beta p^2 /2}$
is the  normalized Maxwellian.\\
(iv) \quad $f_N = 0$ on $\Gamma_N \setminus \Gamma_N^\ast$ and $f_N$
is continuous along trajectories of $T_t^{(N)}$ on $\Gamma_N^\ast$,
cf.
(\ref{7}) and (\ref{8}).\\
(v) \quad The time evolved measure $P_t$ has a density $f_N(t)$
given by $ f_N(t)= 0$ on $\Gamma_N \setminus \Gamma_N^\ast$ and
\begin{equation}\label{14b}
f_N (x_1,\ldots, x_N,t) = f_N (T_{-t}^{(N)} (x_1,\ldots,x_N))
\end{equation}
for $(x_1, \ldots, x_N) \in \Gamma_N^\ast$. The canonical
equilibrium measure is denoted by $P_{\mathrm{eq}}$.

To avoid confusion we remark that identities are always understood
pointwise. If they hold only a.s., we state this explicitly. Often
we will work with densities of measures. As in the case of $f_N(t)$
and of $\rho_n(t)$ below we will choose then a specific version.

Because of a definite number of particles, $N$, the correlation
functions are, up to a multiplicative factor, just the marginal
measures. We fix a particular version of these measures by
\begin{equation}\label{14c}
\rho_n(x_1,\ldots, x_n, t) =N \ldots (N-n+1) \int_{\Gamma_{N-n}
(x_1,\ldots, x_n)} dx_{n+1}\ldots dx_n f_N (x_1,\ldots, x_N, t)\,.
\end{equation}
Note that $\rho_n(t) = 0~\mbox{on}~ \Gamma_n \setminus
\Gamma_n^\ast$ by our definition of $ \Gamma_{N-n} (x_1,\ldots,
x_n)$. Let $\Delta \subset \Gamma_n$ be a Borel set. We remove a set
of Lebesgue measure zero to guarantee that $\Delta \subset
\Gamma_n^\ast$. Then because of the hard core exclusion and by
symmetry
\begin{eqnarray}\label{14d}
&&\hspace{0pt}\lefteqn{\int_{\Delta} dx_1\ldots dx_n \rho_n
(x_1,\ldots,
x_n,t)}\nonumber\\
&&\hspace{80pt}= N\ldots(N-n+1) P\{(x_1(t), \dots, x_n (t)) \in
\Delta\}\,,
\end{eqnarray}
where for $j=1,\ldots,N$ we set
\begin{equation}\label {14e}
x_j (t\pm, x) = (T_{t\pm}^{(N)} \,x)_j \,.
\end{equation}
The probability in (\ref{14d}) is independent of whether the limit
is taken from the future of from the past.

To avoid an overburdened language it is convenient to set
\begin{equation}\label{14f}
t > 0\,,\nonumber
\end{equation}
which we do from now on. This is no restriction, of course.

Let $\tau_m \ge 0$, $m=1,2,\ldots,$ be the time of the $m$--th
collision between the set of particles with labels $1,\ldots,n$
and the set of particles with labels $n+1,\ldots,N.$ If there are
simultaneous collisions between the two groups of particles, then
they are ordered according to the label in the first group.
\begin{Proposition}\label{prop1}
The following idendity holds for all Borel sets $\Delta \subset
\Gamma_n^\ast $, for all $ n=1,2,\ldots,N$,
\begin{eqnarray}\label{15}
&&\hspace{-30pt}\int_\Delta dx_1 \dots dx_n  \rho_n(x_1,\ldots,
x_n,t)\nonumber\\
 &&\hspace{-20pt}=\int_\Delta dx_1 \dots dx_n 
\rho_n(T_{-t}^{(n)} (x_1,\ldots,
x_n))\nonumber\\
&&\hspace{-10pt}+\sum_{m=1}^\infty N \dots (N-n+1)
P\big\{(x_1(\tau_m+),\ldots, x_n(\tau_m+))\in
T_{\tau_m-t+}^{(n)} \Delta, \tau_m \le t\big\}\nonumber\\
&&\hspace{-10pt}-\sum_{m=1}^\infty N \dots (N-n+1)   P\big\{
(x_1(\tau_m -),\ldots, x_n(\tau_m-))\in T_{\tau_m-t-}^{(n)}\Delta,
\tau_m \le t\big\}\,.
\end{eqnarray}
\end{Proposition}
\begin{Proof}
 We use inclusion - exclusion to obtain
\begin{eqnarray}\label{16}
&&\hspace{-10pt} P\big\{(x_1(t), \ldots, x_n(t)) \in \Delta \big\}\nonumber\\
&&\hspace{0pt}=P \big\{(x_1(t-), \ldots, x_n(t-)) \in \Delta, \tau_1 \le t \big\}\nonumber\\
&&\hspace{10pt} + P\big\{(x_1(t+), \ldots, x_n(t+)) \in \Delta,
\tau_1>t\big\}\nonumber\\
&&\hspace{0pt}=\sum_{m=1}^\infty \, P \big\{(x_1(t-), \ldots,
x_n(t-)) \in \Delta, \tau_{m+1} > t,
\tau_m \le t\big\}\nonumber\\
&&\hspace{10pt} + P \big\{(x_1, \ldots, x_n) \in T_{-t-}^{(n)} \Delta, \tau_1 >t\big\}\nonumber\\
&&\hspace{0pt}=\sum_{m=1}^\infty \, P \big\{(x_1(\tau_m+), \ldots,
x_n(\tau_m+)) \in T_{\tau_m-t+}^{(n)}
\Delta, \tau_{m+1} > t,\tau_m \le t \big\}\nonumber\\
&&\hspace{10pt} + P \big\{(x_1, \ldots, x_n) \in T_{-t}^{(n)} \Delta
\big\}-P\big\{(x_1,\ldots,x_n) \in
T_{-t-}^{(n)} \Delta, \tau_1 \le t\big\}\nonumber\\
&&\hspace{0pt}=\sum_{m=1}^\infty \, P \big\{(x_1(\tau_m +), \ldots,
x_n (\tau_m +))\in T_{\tau_{m-t+}}^{(n)}
\Delta, \tau_m \le t\big\} \nonumber \\
&&\hspace{10pt}-\sum_{m=1}^\infty \, P \big\{(x_1(\tau_m +), \ldots,
x_n(\tau_m +)) \in T_{\tau_{m-t+}}^{(n)}
\Delta, \tau_{m+1}\le t\big\}   \mbox{ \hspace{4pt}($\ast$)} \nonumber \\
&&\hspace{10pt} + P\big\{(x_1,\ldots, x_n) \in T_{-t}^{(n)} \Delta\big\}\nonumber\\
&&\hspace{10pt} - \sum_{m=1}^\infty \, P \big\{(x_1(\tau_m -),
\ldots, x_n(\tau_m -)) \in
T_{\tau_{m-t-}}^{(n)} \Delta, \tau_m \le t\big\}\nonumber\\
&&\hspace{10pt} + \sum_{m=1}^\infty \, P \big\{(x_1(\tau_{m+1} -),
\ldots, x_n(\tau_{m +1}-)) \in
 T_{\tau_{m+1}-t-}^{(n)} \Delta, \tau_{m+1}\le
 t\big\}\,.\mbox{ \hspace{4pt}($\ast\ast$)}
\end{eqnarray}
To justify (\ref{16}) we need an integrable bound. Clearly, the
$m$-th term is bounded by $c P_{\mathrm{eq}} \{\tau_m \le t\}$. We
will show in Lemma \ref{prop2} below that this bound is summable.

In (\ref{16}) ($\ast$) and ($\ast\ast$) cancel each other because as
sets
\begin{eqnarray}\label{20}
&&\hspace{-30pt}\big\{(x_1,\ldots, x_N ) \in  \Gamma_N^\ast
\big|(x_1(\tau_m + ) , \ldots, x_n(\tau_m +))
\in T_{\tau_m{-t +}}^{(n)}\Delta, \tau_{m+1}\le t\big\}\nonumber\\
&&\hspace{-25pt}= \big\{(x_1,\ldots, x_N) \in  \Gamma_N^\ast
\big|(x_1(\tau_{m+1}-) ,\ldots, x_n (\tau_{m+1}-)) \in
 T_{\tau_{m+1}-t-}^{(n)}\Delta, \tau_{m+1} \le t\big\}\,.
\end{eqnarray}
\end{Proof}
\begin{Lemma}\label{prop2}
Let $ \tau_m \ge 0$, $m=1,2,\dots,$ be
the time of the $m$-th collision for the system of $N$ hard spheres
(collisions with the wall are not counted). Then
\begin{eqnarray}\label{21}
&&\hspace{-20pt}\sum_{m=1}^\infty P_{\mathrm{eq}} \{ \tau_m \le t \}\nonumber\\
&&\hspace{-10pt}= t \int dq_1 dp_1 \int dp_2
\int_{\Omega_{1-}(x_1,p_2)} d\hat\omega ~a^2 \hat\omega
\cdot(p_1-p_2) \rho_{\mathrm{eq},2} (q_1, p_1, q_1 + a\hat\omega,
p_2)\,.
\end{eqnarray}
\end{Lemma}
\begin{Proof}
 We think of the hard sphere dynamics as
a flow under a function (special flow), cf. \cite{4} for this
construction in our context, and we prove Lemma \ref{prop2} first for this
case.

Let $B$ be the base and $T: B\to B$ be an invertible map which
preserves the finite measure $\mu$. Let $h: B\to \mathbb R_+$
be the ceiling function. We assume that $h$ is integrable. The phase
space is then $\Gamma = \{ x \in B, y \in {\mathbb R}_+\, |\, 0 \leq
y \leq h(x)\}.$ The flow $T_t$ is constructed piecewise in the
following way: $T_t: (x,y) \mapsto (x,y+t)$ until the first time for
which $y+t=h(x).$ Then $(x,h(x)) \mapsto (T x,0).$ We refer to this
transformation as a collision. The construction is then continued
into the future and the past. The measure $\mu (dx) \times dy =
P_{\mathrm{eq}}$ is invariant under $T_t.$ Let $\tau_m \ge 0$,
$m=1,2,\ldots,$ be the time of the $m$-th collision. Then we claim
that
\begin{equation}\label{22}
\sum_{m=1}^\infty P_{\mathrm{eq}} \{ \tau_m \le t\} = t \mu(B)\,.
\end{equation}

Since $h>0, \sum_{j=0}^\infty h (T^{-j} x) = \infty$ $\mu(dx)$ a.s.
by the Poincar\'e recurrence theorem. Therefore
\begin{equation}\label{23}
B_k=\big\{x\in B \big| \sum_{j=0}^{k-2} h(T^{-j} x) \le t,
\sum_{j=0}^{k-1} h(T^{-j} x) > t\big\},
\end{equation}
$k=1,2,\ldots,$ forms a partition of $B$. Then
\begin{eqnarray}\label{24}
&&\hspace{-10pt}\sum_{m=1}^\infty P\{ \tau_m \le t\} = t \mu(B_1)+
\sum_{k=2}^\infty
\int_{B_k} \mu(dx)h(x)\nonumber\\
&&\hspace{10pt}+\sum_{m=2}^\infty \Big\{ \int_{B_m} \mu (dx) \big(t-
\sum_{j=0}^{m-2} h(T^{-j} x)\big)+
\sum_{k=m+1}^\infty  \int_{B_k} \mu (dx)  h(T^{-k+2} x)\Big\}\nonumber\\
&&\hspace{0pt}= t \sum_{k=1}^\infty \mu(B_k) = t \mu(B)\,.
\end{eqnarray}

For hard spheres the base consists of configurations with outgoing
momenta and is defined by
\begin{eqnarray} \label{25}
&&\hspace{-10pt}B=\big\{( x_1,\ldots, x_N )\in \Gamma_N^\ast\big| ~
\mbox{there exists a pair}~
(i,j), i\not= j,\nonumber\\
&&\hspace{50pt}\mbox{such that}\quad q_j = q_i+a \hat\omega,
(p_j-p_i)\cdot \hat \omega > 0\big\}.
\end{eqnarray}
The ceiling function is defined as the time until the next
collision (not counting collisions with the wall). The equilibrium
measure induces on $B$ the invariant surface measure
\begin{equation}\label{26}
\big\{ \sum_{i\not= j=1}^N a^2 dq_i d\hat\omega ~\hat\omega \cdot
(p_j-p_i) \prod_{k=1, k\not=i,j}^N dq_k \big\} \prod_{k=1}^N h_\beta
(p_k) dp_k\,.
\end{equation}
Its total weight is given by (\ref{21}).\medskip
\end{Proof}

We want to express (\ref{15}) in terms of correlation functions. For
this purpose we first have to show some regularity of these
functions.
\begin{Lemma}\label{prop3} 
Under our assumptions on $P$, for
every $s \in {\mathbb R}$, $\rho_n(s) =0$ on $ \Gamma_n \setminus
\Gamma_n^\ast$ and $ \rho_n(s)$ is continuous along trajectories of
$ T_t^{(n)}$ on $\Gamma_n^\ast.$
\end{Lemma}
\begin{Proof} Since, by assumptions (iv) and (v), $f_N(s)$
has the same continuity properties as $f_N(0),$ we may set
$s=0.$

To simplify notation we abbreviate $x=(x_1, \ldots, x_n), y =
(x_{n+1},\ldots, x_N)$ and we set $x(t\pm, x)= T_{t\pm}^{(n)}x$. For
$x \in \Gamma_n^\ast$ let $\Lambda(x,t) \subset \Lambda$ be spatial
region traced out by the particles' motion $x(s), 0 \le s \le  t$,
with initial condition $x$. We set $\Lambda(x,0)=\Lambda(x).$
Correspondingly we define $\Lambda(y,t) \subset \Lambda$ for $y \in
\Gamma_{N-n}^\ast.$ Then for $x \in \Gamma_n^\ast$ let
\begin{equation}\label{28}
\Gamma_{N-n} (x,t) = \big\{ y \in \Gamma_{N-n}^\ast |
\Lambda(x,t)\cap \Lambda (y,t)=\emptyset \big\}.
\end{equation}
We have $\Gamma_{N-n}(x)= \Gamma_{N-n} (x,0)$ up to a set of
$dy$-measure zero. For $x \in \Gamma_n^\ast$ we define the flow
$T_t^{(x)}$ on $\Gamma_{N-n}(x)^\ast$ of $N-n$ particles in the
spatial region $\Lambda \setminus \Lambda(x)$. Here the $^\ast$
indicates again that we remove from $\Gamma_{N-n}(x)$ a set of
Lebesgue measure zero on which the flow remains undefined.

With these definitions, for $x \in \Gamma_n^\ast$,
 \begin{eqnarray}\label{29}
&&\hspace{-20pt} \rho_n(T_{t\pm}^{(n)} x) =
\int_{\Gamma_{N-n}(x(t))} dy N \ldots (N - n +1) f_N
 (T_{t\pm}^{(n)} x,y) \nonumber\\
&&\hspace{32pt}= \int_{\Gamma_{N-n}(x(t))} dy N \ldots (N-n+1) f_N
(T_{t\pm}^{(n)} x, T_t^{(x(t))} y)
 \end{eqnarray}
 by Liouville's theorem for the map $T_t^{(x(t))}$ for fixed
 $t$.

We choose now a $\tau$ such that $0<t\le\tau$. For $x
\in\Gamma_n^\ast$ and $y \in \Gamma_{N-n}(x,\tau)$ the
``$x$''-particles and the
 ``$y$''-particles do not interact
 during the time interval $[0,\tau]$. We then have two
 possibilities:\medskip\\
 (1) The time evolution exists into the future and the past, i.e.
 $(x,y)\in
 \Gamma_N^\ast$. In this case $(T_{t\pm}^{(n)} x, T_{t\pm}^{(x(t))} y) = T_{t\pm}^{(N)}
 (x,y)$. We denote the set of such $y^\prime$s by $\widehat\Gamma_{N-n}
 (x,\tau).$\medskip\\
 (2) The time evolution does not exist, i.e. $(x,y) \notin \Gamma_N^\ast.$
 In this case, by assumption (v),
\begin{equation}
 f_N( T_t^{(n)} x, T_t^{(x(t))} y) = 0 \quad \mbox{for} \quad 0 \le t\le
 \tau.
\end{equation}
 Therefore, for every $x\in \Gamma_n^\ast$,
 \begin{eqnarray}\label{30}
 &&\hspace{-30pt}| \rho_n(T_{t\pm}^{(n)} x) - \rho_n(x)|\nonumber\\
 &&\hspace{-20pt}= N \ldots (N-n+1)\Big| \int_{\widehat\Gamma_{N-n}(x,\tau)} dy f_N
 (T_t^{(N)} (x,y))\nonumber\\
 &&\hspace{-15pt}+ \int_{\Gamma_{N-n}(x(\tau))\setminus
 \Gamma_{N-n}(x,\tau)} dy f_N(T_{t\pm}^{(n)} x, T_t^{(x(t))}
 y) - \int_{\Gamma_{N-n}(x)} dy f_N(x,y)\Big|\nonumber\\
 &&\hspace{-20pt} \le N \ldots (N-n+1)\Big\{\int_{\widehat\Gamma_{N-n}(x,\tau)}
 dy | f_N (T_t^{(n)} (x,y))-f_N(x,y)|\nonumber\\
 &&\hspace{-15pt}+ c \int_{\Gamma_{N-n}(x(\tau))\setminus\Gamma_{N-n}(x,\tau)} dy 
 f_{\mathrm{eq},N} (x,y)+ c \int_{\Gamma_{N-n}(x)\setminus \Gamma_{N-n} (x,\tau)}
 dy  f_{\mathrm{eq},N} (x,y) \Big\}\,.\label{31}
\end{eqnarray}
The last two terms are bounded by \textit{const}. $\tau$. For fixed
$\tau$ the first term vanishes in the limit $t \to 0$.
 This follows from dominated convergence and our assumption (iv).
\end{Proof}
\begin{Lemma}\label{prop4}
Under our assumptions on $P$, for every  $(x_1,\ldots, x_n) \in
\Gamma_n^\ast$ the map $t \mapsto \rho_n (x_1,\ldots, x_n,t)$ is
continuous, i.e.
\begin{equation}\label{32}
\lim_{t\to 0} ~\rho_n (x_1,\ldots, x_n,s+t) = \rho_n (x_1,\ldots,
x_n,s)\,.
\end{equation}
\end{Lemma}
\begin{Proof}
We use the same notation as in the proof of Lemma \ref{prop3}. Since, by
assumptions (iv) and (v), $f_N(s)$ has the same continuity
properties as $f_N(0)$, we may set $s=0$.

For every $x\in\Gamma_n^\ast$ we have
\begin{eqnarray} \label{33}
&&\hspace{-10pt}\rho_n(x,t) - \rho_n(x)\nonumber\\
&&\hspace{10pt}= \int_{\{y|(x,y)\in\Gamma_N^\ast\}} dy f_N(x,y,t)
-\int_{\{y|(x,y)\in\Gamma_N^\ast\}}
dyf_N(x,y)\nonumber\\
&&\hspace{10pt}=\int_{\{y|(x,y)\in\Gamma_N^\ast\}} dy 
(f_N(T_{-t}^{(N)}(x,y))- f_N (x,y))\,.
\end{eqnarray}
The claim follows then by dominated convergence from assumption
(iv).
\end{Proof}
\begin{Proposition}\label{prop5}
The following identity holds for every Borel set  $\Delta \subset
\Gamma_n^\ast$, $n=1,2, \ldots, N$,
\begin{eqnarray}\label{34}
&&\hspace{-10pt}\int_\Delta d x_1 \ldots dx_n \rho_n (x_1,\ldots,
x_n,t)\nonumber\\
&&\hspace{0pt}=\int_\Delta d x_1 \ldots dx_n \rho_n
(T_{-t}^{(n)}(x_1,\ldots,
x_n))\nonumber\\
&&\hspace{10pt}+ \sum_{j=1}^n \int_0^t ds \int_\Delta d x_1 \ldots
dx_n \,[C_{j,n+1} \rho_{n+1} (s)] \,(T_{-t+s}^{(n)}(x_1,\ldots,
x_n))\,.
\end{eqnarray}
Here the collision operator is defined by
\begin{eqnarray} \label{35}
&&\hspace{-10pt}(C_{j,n+1} \rho_{n+1}(s))(x_1,\ldots, x_n)\\
&&\hspace{0pt}= a^2 \int d p_{n+1} \int_{\Omega(x_1,\ldots,x_n,
p_{n+1})}    d\hat{\omega}\hat{\omega}\cdot (p_{n+1}
- p_j) \rho_{n+1} (x_1,\ldots,x_n, q_j + a \hat{\omega},p_{n+1}, s)\,.\nonumber
\end{eqnarray}
\end{Proposition}
\begin{Proof}
We consider the third term of (\ref{15}), cf. Proposition \ref{prop1}. For $0
\le s < t$ we want to compute the limit as $\varepsilon \to 0$ of
\begin{eqnarray} \label{37}
&&\hspace{-30pt}\lefteqn{\sum_{m=1}^\infty \frac{1}{\varepsilon} P\big\{(x_1 (\tau_m-),  \ldots,
x_n(\tau_m-)) \in T_{\tau_m-t-}^{(n)}\Delta,\tau_m \in
[s,s+ \varepsilon]\big\}}\nonumber\\
&&\hspace{-20pt}=\sum_{k=1}^\infty k \frac{1}{\varepsilon} P_s\big\{\mbox{particles with labels
$1,\ldots, n$ collide exactly $k$ times with}\nonumber\\
&&\hspace{20pt}\mbox{particles with labels $n+1,\ldots,N$
during the time interval $[0,\varepsilon]$}, \nonumber\\
&&\hspace{20pt}\mbox{at the times $\tau$ of collision }
 (x_1(\tau-),\ldots,x_n( \tau-)) \in
T_{s+\tau-t-}^{(n)}\Delta\big\}\,.
\end{eqnarray}
Here $P_s$ is the measure $P$ evolved to time $s$. By assumption
(iii) the sum for $k \ge 2$ is bounded by
\begin{equation} \label{38}
\sum\limits_{k=2}^\infty k \frac{c} {\varepsilon}
P_{\mathrm{eq}}\big\{\mbox{particles have exactly $k$
 collisions during the time interval}~ [0,\varepsilon]\big\}.
\end{equation}
By the same argument as in Lemma \ref{prop2}
\begin{eqnarray} \label{39}
&&\hspace{-40pt}\lim_{\varepsilon \to 0} \frac{1}{\varepsilon} \,
P_{\mathrm{eq}}
\big\{\mbox{particles have exactly one collision during the time interval} ~[0,\varepsilon]\big \}\nonumber\\
&&\hspace{-20pt}=\int dq_1 dp_1\int dp_2 \int_{\Omega_{1-}(x_1,p_2)}
d \hat{\omega} ~a^2\hat\omega\cdot(p_1-p_2)
    \rho_{\mathrm{eq},2} (q_1, p_1, q_1 + a \hat\omega, p_2)
\end{eqnarray}
and according to Lemma \ref{prop2}
\begin{eqnarray} \label{40}
&&\hspace{-40pt}\lefteqn{\sum\limits_{k=1}^\infty k P_{\mathrm{eq}}
\big\{\mbox{particles have exactly $k$
collisions during the time interval}~ [0,\varepsilon]\big\}}\nonumber\\
&&\hspace{-20pt}= \varepsilon \int dq_1 dp_1\int dp_2
\int_{\Omega_{1-}(x_1, p_2)} d \hat\omega~
a^2\hat\omega\cdot(p_1-p_2)
    \rho_{\mathrm{eq},2} (q_1, p_1, q_1 + a \hat\omega, p_2)\,.
\end{eqnarray}
Therefore in the limit $\varepsilon \to 0$ the expression in (\ref{38}) vanishes.

We are left with the term $k=1$ of (\ref{37}).
Let us label the particle at the collision with $n+1.$ Then we
have to compute the limit $\varepsilon \to 0$ of
\begin{eqnarray}\label{41}
&&\hspace{-30pt}\lefteqn{(N-n) \frac{1}{\varepsilon}P_s \big\{ \mbox{particles with label $1,
 \ldots, n$ collide exactly once with particle $n+1$}} \nonumber\\
 && \mbox{and do not collide with particles with
 label $n+2, \ldots, N$ during the time} \nonumber\\
 && \mbox{interval $[0,\varepsilon]$, at the time $\tau$ of
 collision }(x_1(\tau-), \ldots, x_n (\tau-)) \in T_{s+\tau-t-}^{(n)}
 \Delta
\big \} \nonumber\\
 &&\hspace{-30pt}=\sum\limits_{j=1}^n (N-n) \frac{1}{\varepsilon} P_s \big\{ \mbox{ the only
 collision during time interval $[0, \varepsilon]$ is} \nonumber\\
 && \mbox{between particle $j$ and particle $n+1$, at time $\tau$ of collision } \nonumber\\
 && (x_1(\tau -),\ldots, x_n(\tau-)) \in T_{s+\tau-t-}^{(n)} \Delta \big\} + \mathcal{O}(\varepsilon)\,.
\end{eqnarray}
The error is bounded by
\begin{equation}\label{41a}
 \frac{c}{\varepsilon} P_{\mathrm{eq}}\big\{\mbox{there is more than one collision during
the time interval}~[0,\varepsilon]\big\}\,,
\end{equation}
which vanishes in the limit  $\varepsilon \to 0$.

Let
\begin{eqnarray}\label{42}
&&\hspace{-50pt} A_j(\varepsilon)= \big\{(x_1, \ldots, x_{n+1}) \in
\Gamma_{n+1}^\ast \big| T_t^{(n+1)}(x_1,
\ldots, x_{n+1})  \mbox{ for } 0 \le t \le \varepsilon\nonumber\\
&&\hspace{0pt}\mbox{has as only collision the one between particles  $j$
and $n+1$},\nonumber\\
&&\hspace{0pt} \mbox{at time $\tau$ of collision } (x_1(\tau-),
\ldots, x_n(\tau-)) \in T_{s+\tau-t-}^{(n)} \Delta \big\}.
\end{eqnarray}
In the definition (\ref{42}) $x_j(t\pm, x) = (T_{t\pm}^{(n+1)}x )_j.$
If the set defined in (\ref{41}) is called $B_j (\varepsilon)$, then
\begin{eqnarray}\label{43}
&&\hspace{-20pt}\lefteqn{(\ref{41})= \sum_{j=1}^n (N-n)
\frac{1}{\varepsilon}
         \int_{B_j(\varepsilon)} dx_1 \ldots dx_N f_N (x_1, \ldots,
    x_N, s)} \nonumber\\
&&\hspace{-20pt}=\sum_{j=1}^n (N-n)
\frac{1}{\varepsilon}\int_{A_j(\varepsilon)}
   dx_1 \ldots dx_{n+1} \int_{\Gamma_{N-n-2} (x_1, \ldots,
   x_{n+1})}dx_{n+2}\ldots dx_N
   f_N (x_1, \ldots, x_N, s) \nonumber\\
&&\hspace{0pt} + \sum_{j=1}^n (N-n)
\frac{1}{\varepsilon}\Big[\int_{B_j(\varepsilon)}
     dx_1 \ldots dx_N f_N (x_1, \ldots, x_N, s)\nonumber\\
&&\hspace{0pt} - \int_{A_j(\varepsilon)} dx_1 \ldots dx_{n+1}
\int_{\Gamma_{N-n-2}
    (x_1, \ldots, x_{n+1})} dx_{n+2} \ldots dx_N f_N (x_1, \ldots, x_N,
    s)\Big]\,.
\end{eqnarray}
The second term is again bounded by (\ref{41a}) and vanishes
therefore in the limit $\varepsilon \to 0$.

Multiplying with the factor $N \ldots (N-n+1)$ of (\ref{15}) we are
left with
\begin{equation}\label{44}
\sum_{j=1}^n  \frac{1}{\varepsilon}\int_{A_j(\varepsilon)}
     dx_1 \ldots dx_{n+1} \rho_{n+1}  (x_1, \ldots, x_{n+1}, s)\,.
\end{equation}
Let $\tau$, $0 \le \tau \le \varepsilon$, be the time of collision.
Then on $A_j (\varepsilon)$
\begin{equation}\label{45}
q_{n+1} = q_j + a \hat \omega + \tau(p_j-p_{n+1})\,.
\end{equation}
We perform this substitution in the integral (\ref{44}). The
change in volume element is
\begin{equation}\label{46}
dq_j dp_j dq_{n+1} dp_{n+1} =a^2 \hat \omega \cdot (p_j-p_{n+1})
dq_j dp_j d \tau d \hat \omega dp_{n+1}\,.
\end{equation}
We flow on $A_j(\varepsilon)$ all $n+1$ coordinates from time 0 to
time $\tau$. Then
\begin{eqnarray}\label{47}
&&\hspace{-20pt}\lefteqn{\sum_{j=1}^n
\frac{1}{\varepsilon}\int_{A_j(\varepsilon)}
     dx_1 \ldots dx_{n+1}  \rho_{n+1}  (x_1, \ldots, x_{n+1}, s)}\nonumber\\
&&\hspace{-5pt}= \sum_{j=1}^n \frac{1}{\varepsilon}\int_0^\varepsilon
   d \tau a^2 \int_{T_{s+ \tau-t-}^{(n)} \Delta}
   dx_1\ldots dx_n
   \int dp_{n+1} \int_{\Omega_{j-} (x_1, \ldots, x_n,
   p_{n+1})} d \hat\omega \hat\omega \cdot (p_j-p_{n+1}) \nonumber\\
&&\hspace{30pt} \times\chi_\Xi (x_1, \ldots, x_n, \hat\omega,
p_{n+1})\rho_{n+1} (T_{-\tau}^{(n+1)} (x_1, \ldots, x_n, q_j +
a \hat\omega, p_{n+1}), s)\nonumber\\
&&\hspace{-5pt}= \sum_{j=1}^n  \frac{1}{\varepsilon}
\int_0^\varepsilon d \tau a^2
    \int_{T_{s+ \tau-t-}^{(n)} \Delta} dx_1\ldots dx_n
    \int dp_{n+1}
   \int_{\Omega_{j-} (x_1, \ldots, x_n,
   p_{n+1})} d \hat\omega \hat\omega \cdot (p_j-p_{n+1}) \nonumber\\
&&\hspace{30pt}\times\rho_{n+1} (T_{-\tau}^{(n+1)} (x_1, \ldots, x_n, q_j
+ a \hat\omega, p_{n+1}), s) + \mathcal{O}(\varepsilon)\,.\label{48}
\end{eqnarray}
In the second integral $\chi_\Xi$ is the indicator function of the
set $\{ x_1, \ldots, x_n, \hat\omega, p_{n+1} |$ $
T_{\tau\prime}^{(n+1)} (x_1, \ldots, x_n, q_j + a \hat\omega,
p_{n+1}) \mbox{ for}~
      -\tau \le \tau^\prime \le \varepsilon- \tau
     ~\mbox{has only one collision} \}$.
[As collisions we always refer to collisions between two particles
and not to collisions with the wall. Therefore in (\ref{45}) we
should actually use the free flow of particles $j$ and $n+1$
separately {\it including} collisions with the wall. After flowing
the $n+1$ coordinates to time $\tau$ we still obtain (\ref{47}).]
The error term is again bounded by (\ref{41a}). Only
$P_{\mathrm{eq}}$ refers now to the equilibrium measure of $n+1$
particles.

To obtain the limit as $\varepsilon \to 0$ of (\ref{48}) we have
to show that the integrand is continuous at $\tau=0$. To see this
we bound as
\begin{eqnarray}\label{49}
&&\hspace{-20pt}\lefteqn{\Big| \int_{T_{s+ \tau-t}^{(n)} \Delta}
dx_1 \ldots dx_n \int dp_{n+1} \int_{\Omega_{j-} (x_1, \ldots, x_n,
   p_{n+1})} d \hat\omega \hat\omega \cdot (p_j-p_{n+1})}\nonumber\\
&&\hspace{60pt}\times\rho_{n+1} (T_{-\tau}^{(n+1)} (x_1, \ldots, x_n, q_j
+
a \hat\omega, p_{n+1}), s)\nonumber\\[1ex]
&&\hspace{-10pt}-\int_{T_{s-t}^{(n)} \Delta} dx_1 \ldots dx_n
   \int dp_{n+1} \int_{\Omega_{j-} (x_1, \ldots, x_n,
   p_{n+1})} d \hat\omega \hat\omega \cdot
   (p_j-p_{n+1})\nonumber\\
&&\hspace{60pt}\times\rho_{n+1}  (x_1, \ldots, x_n, q_j + a \hat\omega,
p_{n+1},
s)\Big|\nonumber\\[1ex]
&&\hspace{-20pt} \le \int_{(T_{s+ \tau-t}^{(n)} \Delta \cup
T_{s-t}^{(n)} \Delta)\setminus {(T^{(n)}_{s+ \tau-t}} \Delta\cap
T_{s-t}^{(n)} \Delta)} dx_1 \ldots dx_n
\int dp_{n+1} \int_{\Omega_{j-}(x_1, \ldots, x_n,p_{n+1})}d \hat\omega\nonumber\\[1ex]
&&\hspace{60pt}\times \hat\omega \cdot (p_j- p_{n+1})
\rho_{n+1}(T_{-\tau}^{(n+1)}(x_1, \ldots, x_n, q_j + a \hat\omega,
p_{n+1}), s)\nonumber\\[1ex]
&&\hspace{0pt}+ \int_{T_{s-t}^{(n)} \Delta} d x_1 \ldots dx_n \int
dp_{n+1}
 \int_{\Omega_{j-}(x_1, \ldots, x_n, p_{n+1})} d\hat
\omega \hat\omega\cdot (p_j-p_{n+1})\\
&&\hspace{-20pt}\times\big| \rho_{n+1} (T_{-\tau}^{(n+1)} (x_1, \ldots,
x_n,q_j + a \hat\omega, p_{n+1}), s) - \rho_{n+1}(x_1,\ldots, x_n,
q_j+ a \hat\omega, p_{n+1}, s)\big|\,.\nonumber
\end{eqnarray}
In the first term we bound $\rho_{n+1}^{(s)}$ by 
\textit{const}.$f_{\mathrm{eq}, n+1}.$ By dominated convergence this term vanishes
then in the limit $\tau \to 0$. In the second term we integrate only
over points such that $(x_1, \ldots, x_n, q_j + a\hat\omega,
p_{n+1}) \in \Gamma_{n+1}^\ast.$ Therefore by Lemma \ref{prop3} the integrand
is continuous in $\tau$ and vanishes as $\tau \to 0.$

Altogether we have shown that the measure
\begin{equation}\label{50}
\sum_{m=1}^\infty N \ldots (N-n+1) P\{ (x_1(\tau_m-), \ldots,
x_n(\tau_m-)) \in T_{\tau_m -t-}^{(n)} \Delta, \tau_m \in ds\}
\end{equation}
is absolutely continuous with respect to the Lebesgue measure and
has a density given by
\begin{eqnarray}\label{51}
&&\hspace{-20pt}\sum_{j=1}^n a^2\int_{T_{s-t}^{(n)}\Delta} dx_1
\ldots dx_n \int dp_{n+1} \int_{\Omega_{j-}(x_1, \ldots, x_n,
p_{n+1})}
d\hat\omega \hat\omega \cdot (p_j-p_{n+1}) \nonumber\\
&&\hspace{100pt}\times\rho_{n+1} (x_1, \ldots, x_n, q_j+a \hat\omega,
p_{n+1}, s)\,.
\end{eqnarray}
We note that by Lemma \ref{prop4} this density is continuous in $s$.

The same argument applied to the second term of (\ref{15}) shows
that
\begin{equation}\label{52}
\sum_{m=1}^\infty N \ldots (N-n+1) P \{ (x_1(\tau_m+), \ldots,
x_n(\tau_m+)) \in T_{\tau_m -t+}^{(n)}\Delta, \tau_m \in ds \}
\end{equation}
has a density given by
\begin{eqnarray}\label{53}
&&\hspace{-18pt}\sum_{j=1}^n a^2\int_{\{(x_1, \ldots, q_j,
p_j^\prime, \ldots,x_n)\in
  T_{s-t}^{(n)}\Delta\}} dx_1  \ldots dx_n\int dp_{n+1}
  \int_{\Omega_{j-}(x_1, \ldots, x_n, p_{n+1})}
  d\hat\omega \hat\omega \cdot (p_j-p_{n+1})\nonumber\\
&&\hspace{80pt}\times\rho_{n+1} (x_1, \ldots, x_n,
  q_j+a \hat\omega, p_{n+1}, s)\nonumber\\[1ex]
&&\hspace{-20pt}=\sum_{j=1}^n a^2\int_{\{(x_1, \ldots,
q_j,p_j^\prime,
  \ldots,x_n) \in T_{s-t}^{(n)}\Delta\}} dx_1 \ldots dq_j dp_j^\prime
  \ldots dx_n \int dp_{n+1}^\prime\\
&&\hspace{-22pt}  \times \int_{\Omega_{j+} (x_1, \ldots, q_j, p_j^\prime, \ldots, x_n,
   p_{n+1}^\prime)} d\hat\omega\hat\omega \cdot (p_{n+1}^\prime-
p_j^\prime)\rho_{n+1}(x_1,\ldots, q_j,p_j^\prime,\ldots, x_n,
q_j+a\hat
   \omega, p_{n+1}^\prime, s),\nonumber
\end{eqnarray}
where we used again Lemma \ref{prop3} which ensures that on the domain of
integration $\rho_{n+1}^{(s)}$ is continuous  through a collision.
We relabel in (\ref{53}) $(p_j^\prime, p_{n+1}^\prime)$ as $(p_j,
p_{n+1})$ and subtract (\ref{51}) from (\ref{53}). Since $\Delta
\subset \Gamma_n^\ast$\,,
\begin{eqnarray}\label{53a}
&&\hspace{0pt}\sum_{j=1}^n \int_0^t ds \int_{T_{s-t}^{(n)} \Delta}
dx_1
\ldots dx_n [C_{j,n+1} \rho_{n+1}(s)] (x_1,\ldots,x_n)\nonumber\\
&&\hspace{0pt}=\sum_{j=1}^n \int_0^t ds \int_\Delta dx_1 \ldots dx_n
[C_{j,n+1} \rho_{n+1}(s)] (T_{s-t}^{(n)} (x_1,\ldots,x_n))\,.
\end{eqnarray}\medskip
\end{Proof}

To obtain the integrated from of the BBGKY hierarchy we have to
iterate (\ref{34}). For this purpose we go back to (\ref{44}). Since
we integrate there over a Borel set of $\Gamma_{n+1}$, we could have
chosen any other version of $\rho_{n+1}(s)$, i.e. any other function
$\tilde\rho_{n+1} (s)$ such that $\rho_{n+1}(s)=\tilde \rho_{n+1}
(s)$ $ dx_1 \ldots d x_{n+1}$ a.s.. We used however certain
properties of $\rho_{n+1} (s)$ in the proof below (\ref{44}).
Therefore, if we want to replace $\rho_{n+1} (s)~\mbox{by}~ \tilde
\rho_{n+1} (s)$, the latter has to satisfy:
\begin{enumerate}
\item[(1)] $\tilde \rho_{n+1} (s) = \rho_{n+1} (s)$ \quad a.s..
\item[(2)] For fixed $s,~ \tilde \rho_{n+1} (s) $ is continuous
            along trajectories of $T_t^{(n+1)}$ on $ \Gamma_{n+1}^\ast.$
\item[(3)] For every $(x_1, \ldots, x_{n+1}) \in \Gamma_{n+1}^\ast$, $ s\mapsto
           \tilde \rho_{n+1}(x_1, \ldots, x_{n+1},s)$ is
           continuous.
\item[(4)] There exist constants $c^\prime, \beta$ such that
\begin{equation}\label{54}
\tilde \rho_{n+1} (s) \le c^\prime f_{\mathrm{eq},n+1}^{(\beta)}\,.
\end{equation}
\end{enumerate}
\begin{Corollary}\label{prop6}
Let $\tilde \rho_{n+1} (s):\Gamma_{n+1} \to \mathbb {R}$ satisfy
the Properties (\ref{1}) to (\ref{4}) given above. Then
\begin{eqnarray}\label{55}
&&\hspace{0pt}\int_\Delta dx_1 \ldots dx_n  \rho_n (x_1, \ldots,
x_n,
   t)\nonumber\\
&&\hspace{0pt}=\int_\Delta dx_1 \ldots dx_n \rho_n (T_{-t}^{(n)}
(x_1, \ldots,
    x_n))\nonumber\\
&&\hspace{10pt}+\sum_{j=0}^n \int_0^t ds \int_\Delta dx_1 \ldots
dx_n [C_{j,n+1}\tilde \rho_{n+1}(s)] (T_{-t+s}^{(n)}
(x_1,\ldots,x_n))\,.
\end{eqnarray}
\end{Corollary}
\begin{Lemma}\label{prop7}
Let $\hat\rho_n(t)$ be defined  by
\begin{eqnarray}\label{55a}
&&\hspace{0pt} \hat\rho_n(x_1, \ldots, x_n,t) = \rho_n(T_{-t}^{(n)}  (x_1, \ldots,
x_n))\nonumber\\
&&\hspace{10pt}+ \sum_{j=1}^n \int_0^t ds [C_{j,n+1} \tilde
\rho_{n+1}(s)] (T_{-t+s+}^{(n)} (x_1,\ldots,x_n))
\end{eqnarray}
for every point $(x_1,\ldots,x_n) \in \Gamma_n^\ast$, where $\tilde
\rho_{n+1}(s)$ satisfies the above Properties (\ref{1}) to
(\ref{4}). Then $\tilde \rho_n (t)$ satisfies also the above four
properties.
\end{Lemma}
We note that $\hat\rho_n (t) = 0$ on $\Gamma_n \setminus
\Gamma_n^\ast$
by the definition of $\rho_n$ and of $C_{j,n+1}$.\medskip\\
\begin{Proof}
Property (\ref{1}) follows from Corollary \ref{prop6}. Property (\ref{3})
follows from Lemma \ref{prop3} (continuity of $\rho_n$ along trajectories of
$T_t^{(n)}$ on $\Gamma_n^\ast)$ and from Property (\ref{4}) of
$\tilde\rho_{n+1} (s)$.

For Property (\ref{4}) we note that by Assumption (iii) on the
initial measure
\begin{equation}\label{56}
\rho_n (t) \le c^\prime f_{\mathrm{eq},n}^{(\beta)}\,.
\end{equation}
Together with the assumed bound on $\tilde\rho_{n+1} (s)$ this
gives a bound of the desired form with some new constants $c^{\prime\prime},
\beta^\prime$.

For Property (\ref{2}) the first term of (\ref{55a}) is continuous
along trajectories by Lemma \ref{prop3}. Therefore we only have to consider
the second term. We have
\begin{eqnarray}\label{57}
&&\hspace{-10pt}\Big|\sum_{j=1}^n \int_0^t ds [C_{j,n+1} \tilde
\rho_{n+1}(s)] (T_{-t+s+\tau +}^{(n)}
   (x_1,\ldots,x_n))\nonumber\\
&&\hspace{0pt}- \sum_{j=1}^n \int_0^t ds [C_{j,n+1}\tilde
\rho_{n+1}(s)] (T_{-t+s+}^{(n)}
    (x_1,\ldots,x_n))\Big | \nonumber\\
&&\hspace{-10pt} \le \sum_{j=1}^n \big | \{\int_0^\tau ds  +
\int_t^{t+\tau} ds
    \} | [C_{j,n+1}\tilde \rho_{n+1}(s-\tau)] (T_{-t+s+}^{(n)}
    (x_1,\ldots,x_n))| \big| \nonumber\\
&&\hspace{0pt}+ \sum_{j=1}^n \int_0^t ds \big| [C_{j,n+1}\tilde
\rho_{n+1}(s-\tau)-
    C_{j,n+1}\tilde \rho_{n+1}(s)] (T_{-t+s+} (x_1, \ldots, x_n))\big|\,.
\end{eqnarray}
By assumption the integrand of the first term is bounded by $c''
 f_{\mathrm{eq},n}^{(\beta')}$. This implies that the first term
 vanishes in the limit $\tau \to 0.$ In the second term we use
 dominated convergence. By the assumed continuity of $s \mapsto \tilde \rho_{n+1}
(x_1,\ldots,x_{n+1},s)$ for $(x_1,\ldots,x_{n+1}) \in
\Gamma_{n+1}^\ast$ the integrand vanishes pointwise in the limit
$\tau \to 0$.\medskip
\end{Proof}

 By Lemma \ref{prop7} we may set in (\ref{55}) $\tilde\rho_{n+1} (s) = \hat\rho_{n+1} (s)$ and obtain
 \begin{eqnarray}\label{58}
 &&\hspace{-20pt}\int_\Delta dx_1 \ldots dx_n \rho_n
 (x_1, \ldots, x_n, t)=\int_\Delta dx_1 \ldots dx_n  \Big[ \rho_n (T_{-t+}^{(n)} (x_1, \ldots, x_n
     ))\\
&&\hspace{-10pt} +\sum_{j_1=1}^n \int_0^t dt_1
     (C_{j_1,n+1} (\rho_{n+1} \circ T_{-t_1 +}^{(n+1)}))(T_{-t+t_1+}^{(n)}
     (x_1, \ldots, x_n))\nonumber\\
 &&\hspace{-10pt} + \sum_{j_1=1}^n \sum_{j_2=1}^{n+1}\int_0^t dt_1 \int_0^{t_1} dt_2
     (C_{j_1,n+1} (C_{j_2, n+2} \tilde\rho_{n+2} (t_2)) \circ  T_{t_2-t_1+}^{(n+1)} ))
     (T_{t_1-t+}^{(n)} (x_1,\ldots,
     x_n))\Big ]\,.\nonumber
 \end{eqnarray}
 We iterate $N-n$ times and obtain
 \begin{Proposition}\label{prop8}
 The following identity holds for every Borel set $\Delta \subset
 \Gamma_n^\ast$, $n=1, \ldots, N$,
 \begin{eqnarray}\label{59}
 &&\hspace{-20pt}\int_\Delta dx_1 \ldots dx_n \rho_n (x_1, \ldots, x_n,t) =\\
 &&\hspace{-10pt}\sum_{m=0}^{N-n} \sum_{j_1=1}^n\cdots \sum_{j_m=1}^{n+m-1} \int_0^t dt_1
      \ldots
    \int_0^{t_{m-1}} dt_m \int_\Delta
   dx_1 \ldots dx_n\nonumber\\
&&\hspace{-10pt} \times(C_{j_1,n+1} \ldots(C_{j_m, n+m} (\rho_{n+m} \circ
T_{t_m+}^{(n+m)})) \circ
T_{t_m-t_{m-1}+}^{(n+m-1)}\ldots)(T_{t_1-t+}^{(n)} (x_1,\ldots,
x_n))\,.\nonumber
\end{eqnarray}
\end{Proposition}

In a way Proposition \ref{prop8} is our final result. There are however two
reasons for reorganizing somewhat the integral (\ref{59}). First
of all we would like to get rid of the continuity assumptions,
i.e. we would like to extend the validity of (\ref{59}) to a more
general class of initial measures. Secondly the Boltzmann--Grad
limit is not quite apparent in the given form of (\ref{59}).

We introduce the notion of a {\it collision history.} I choose this
name to distinguish it from a sequence of real collisions of the
$N$-particle system. One should remember that the correlation
functions are averaged quantities. The correspondence between
collision histories and sequences of real collisions is only very
indirect.

A collision history is specified by the following list:
\begin{enumerate}
\item[(a)] $n\in \mathbb N$,
\item[(b)] $m \in {\mathbb N} \cup \{0\}$,
\end{enumerate}
[If, as assumed so far, the number of particles equals $N,$ then $1
\le n\le N$ and $0 \le m \le N-n.$]
\begin{enumerate}
\item[(c)] $(x_1, \ldots, x_n) \in \Gamma_n^\ast$,
\item[(d)] $(t_1, \ldots, t_m) \in {\mathbb R}^m$
\end{enumerate}
with the constraint  $0 \le t_m \le \ldots \le t_1 \le t$, $ m \ge
1$,
\begin{enumerate}
\item[(e)] $(j_1, \ldots, j_m) \in {\mathbb N}^m$
\end{enumerate}
with the constraint $1 \le j_1 \le n, \ldots, 1 \le j_m \le n+m-1$,
$m \ge 1$,
\begin{enumerate}
\item[(f)] $(\hat p_1, \ldots, \hat p_m) \in {\mathbb R}^{3m}$, $ m \ge
1$,
\item[(g)] $(\hat\omega_1, \ldots, \hat\omega_m) \in (S^2)^m$, $m \ge 1,$
\end{enumerate}
with a complicated constraint depending on $n,m, x_1, \ldots, x_n,
t_1 \ldots, t_m, j_1, \ldots, j_m,$ $ \hat p_1, \ldots, \hat p_m$
which is defined below. For future convenience we abbreviate a
collision history as $(x_1, \ldots, x_n, \delta)$, where $\delta$
stands for $(m, t_1, \ldots, t_m, j_1, \ldots, j_m, \hat p_1,
\ldots, \hat p_m,$ $\hat\omega _1, \ldots, \hat\omega_m) $.

Given the collision history $(x_1, \ldots, x_n, \delta)$ we
construct an evolution of particles in the following way. We choose
$n$ particles at $(x_1, \ldots, x_n) \in \Gamma_n^\ast $. We
consider this as the phase point at time $t$, $(x_1, \ldots, x_n)
\equiv (x_1 (t), \ldots, x_n(t)) $ and we evolve backwards in time
up to $t=0.$ We evolve the phase point $(x_1(t), \ldots, x_n (t))$
to $T_{-t+t_1+}^{(n)} (x_1(t),\ldots, x_n (t)) \equiv
(x_1(t_1),\ldots, x_n (t_1))$. We add a particle with label $n+1$ at
$q_{j_1} (t_1) + a \hat\omega_1$ with momentum $\hat p_1$. We
require that  $\hat\omega_1 \in \Omega_{j_1} (x_1 (t_1),\ldots, x_n
(t_1), \hat p_{n+1})$, i.e. $(x_1 (t_1),\ldots, x_n (t_1), q_{j_1
}(t_1) + a \hat\omega, \hat p_1) \in \Gamma_{n+1}^\ast .$ We call
this new phase point of $n+1$ particles $(x_1(t_1),\ldots,x_{n+1}
(t_1))$ and evolve it to 
 $T_{-t_1+ t_2+}^{(n+1)} (x_1(t_1),\ldots, x_{n+1} (t_1)) \equiv
 (x_1 (t_2),\ldots, x_{n+1} (t_2))$.  We add a particle with label $n+2$
 at $q_{j_2} (t_2) + a \hat\omega_2$ with momentum $\hat p_2$.
 We require that $\hat\omega_2 \in \Omega_{j_2} (x_1(t_2),\ldots, x_{n+1} (t_2),
  \hat p_{n+2})$. We call this new phase point of $n+2$ particles $(x_1(t_2),
  \ldots, x_{n+2} (t_2))$. The final step is to evolve $(x_1 (t_m), \ldots, x_{n+m}
  (t_m)) \in \Gamma_{n+m}^\ast$ to\\
   $T_{-t_m+}^{(n+m)} (x_1 (t_m), \ldots, x_{n+m} (t_m)) \equiv (x_1(0),\ldots, x_{n+m}
   (0))$. If $m=0$, then we only evolve $(x_1(t),\ldots, x_n (t)) \equiv (x_1,\ldots, x_n)$
to $T_{-t+}^{(n+m)} (x_1 (t), \ldots, x_n (t)) \equiv (x_1(0), \ldots, x_n
(0))$.
To make the dependence on $x_1, \ldots, x_n, \delta$ explicit we
write $x_k (s, x_1, \ldots, x_n, \delta)$ for $0 \le s \le t$, in
particular $x_k(x_1, \ldots, x_n, \delta) \equiv x_k(0)$, $k=1,
\ldots, n+m$.

Let
\begin{equation}\label{60}
\Delta(x_1, \ldots, x_n; [0,t])\nonumber
\end{equation}
be the space of all collision histories for given $n$, starting
configuration $(x_1, \ldots, x_n) \in \Gamma_n^\ast$ and time span
$[0,t]$. $\Delta(x_1, \ldots, x_n;[0,t])$ is a subset of
\begin{equation}\label{60a}
\bigcup\limits_{m \ge 0}\bigcup\limits_{j_1=1}^n \ldots
\bigcup\limits_{j_m=1}^{n+m-1} ({\mathbb R} \times {\mathbb R}^3
\times S^2)^m\nonumber
\end{equation}
defined by the above construction. We define a measure $d\delta$ on
$\Delta( x_1, \ldots, x_n;~ [0,t])\!:d \delta$ is the counting
measure with respect to the discrete indices $m,j_1, \ldots, j_m$
and the Lebesgue measure otherwise.

Given a collision history $(x_1, \ldots, x_n, \delta)$ we define the
weight function by
\begin{eqnarray}\label{61}
W(x_1, \ldots, x_n, \delta)= \prod_{k=1}^m \{ a^2 \hat\omega_k \cdot
(\hat p_k - p_{j_k}(t_k, x_1, \ldots, x_n, \delta))\}\,.
\end{eqnarray}
\begin{Lemma}\label{prop9}
The following identity holds for every Borel set $\Delta \subset
\Gamma_n^\ast$, $n \in \mathbb N$,
\begin{eqnarray}\label{62}
&&\hspace{-20pt}\int_\Delta dx_1 \ldots dx_n \rho_n  (x_1, \ldots,
x_n, t) = \int_\Delta dx_1 \ldots dx_n
\int_{\Delta(x_1, \ldots, x_n; [0,t])} d \delta\\
&&\hspace{0pt}\times W (x_1, \ldots, x_n, \delta)
\rho_{n+m(\delta)}(x_1(x_1, \ldots, x_n,\delta), \ldots,
x_{n+m(\delta)} (x_1, \ldots, x_n,\delta))\,.\nonumber
\end{eqnarray}
\end{Lemma}
\begin{Proof}
We write out (\ref{59}) using the definition of $C_{j,n+1}$. From
Assumption (iii) it follows that
\begin{equation}\label{63}
\rho_{n+m} \le c^{\prime\prime} f_{\mathrm{eq},n+m}\,.
\end{equation}
(The simpler part of) Lanford's estimate on the uniform, in $a^2 N,$
convergence of the BBGKY hierarchy shows that
\begin{eqnarray}\label{64}
&&\hspace{-20pt}\int_\Delta dx_1 \ldots dx_n \int_{\Delta (x,\ldots,
x_n;[0,t])} d \delta | W(x_1,\ldots, x_n, \delta)|
\nonumber\\
&&\hspace{20pt}\times f_{\mathrm{eq},n+m(\delta)} ^{(\beta)}
(x_1(x_1,\ldots, x_n, \delta),
\ldots, x_{n+m(\delta)} (x_1,\ldots, x_n,\delta))\nonumber\\
&&\hspace{0pt} \le c^{\prime\prime\prime} \int_\Delta dx_1 \ldots
dx_n f_{\mathrm{eq},n}^{(\beta')} (x_1,\ldots, x_n)
\end{eqnarray}
with $\beta^\prime < \beta$ for all $t$. The details of this
estimate can be found in F. King's thesis \cite{5}. Therefore the
integrations in (\ref{59}) may be interchanged freely.\medskip
\end{Proof}

In the form (\ref{62}) we can extend our identity to a more
general class of initial measures. In particular we will remove
the restriction of a definite number of particles.

Let $\Gamma$ be the grand canonical phase space,
\begin{equation}
\Gamma = \bigcup_{n\ge 0} \Gamma_n .
\end{equation}
The grand  canonical equilibrium measure with inverse temperature $\beta > 0$
and fugacity $z > 0$ is defined by
\begin{equation}\label{66}
f_{\mathrm{eq},n}^{(z,\beta)}(x_1,\ldots, x_n) \frac{1}{n!}
dx_1\ldots
  dx_n = \frac{1}{Z} \prod\limits_{j=1}^n \{ z h_\beta (p_j) \}
   \frac{1}{n!} dx_1\ldots  dx_n
\end{equation}
on $\Gamma_n, n= 0, 1,\ldots,$ where $Z$ is the normalization
constant. Let $\mathcal C$ be the class of functions $f:\Gamma \to
{\mathbb R}$ such that
\begin{description}
\item (i)  $f_n$ is measurable,
\item (ii)  $f_n$ is symmetric in the particle labels,
\item (iii)  there exist positive constants $M,z,\beta$ such that
\end{description}
\begin{equation}\label{67}
| f_n(x_1,\ldots, x_n) |\le M
f_{\mathrm{eq},n}^{(z,\beta)}(x_1,\ldots, x_n)
\end{equation}
for all $(x_1,\ldots, x_n)\in \Gamma_n$, $n=0,1,\ldots$.

Note that actually $\Gamma_n= \emptyset$ for sufficiently  large $n$
because of the hard core exclusion.

Given $f \in {\mathcal C}$ we define the ``correlation function
vector'' $\rho: \Gamma \to {\mathbb R}$ by
\begin{equation}\label{68}
\rho_n(x_1, \ldots, x_n)=  \sum_{m=0}^\infty \frac{1}{m!}
\int_{\Gamma_m(x_1, \ldots, x_n)}
dx_{n+1} \ldots dx_{n+m} f_{n+m} (x_1, \ldots, x_{n+m}) \\
\end{equation}
 for all $(x_1,\ldots, x_n)\in \Gamma_n$, $n=0,1, \ldots$.
\begin{Lemma}\label{prop10}
Let ${\mathcal F}: {\mathcal C} \to \mathcal C$ be the map defined
by (\ref{68}). Then ${\mathcal F}$ is one--to--one and onto.
\end{Lemma}
\begin{Proof}
We have by assumption (iii)
\begin{eqnarray}
&&\hspace{-10pt}| \rho_n (x_1, \ldots, x_n) | \le M
\sum_{m=0}^\infty \frac{1}{m!} \int  dx_{n+1} \ldots  dx_{n+m}
f_{\mathrm{eq},n+m}^{(z,\beta)} (x_1,
\ldots, x_{n+m})\nonumber\\
&&\hspace{71pt}\le  M e^{|\Lambda| z} f_{\mathrm{eq},n}^{(z,\beta)}
(x_1, \ldots, x_n)\,.
\end{eqnarray}
The inverse map is given by
\begin{equation}\label{70}
f_n (x_1, \ldots, x_n)= \sum_{m=0}^\infty \frac{(-1)^m}{m!} \int
dx_{n+1} \ldots dx_{n+m} \rho_{n+m}(x_1, \ldots, x_{n+m})\,.
\end{equation}
\end{Proof}

Let $P$ be a signed measure on $\Gamma$ with density $f$. Then the
time evolved measure $P_t$ has the density $f(t)$ with one version
given by
\begin{equation}\label{71}
f_n (x_1, \ldots, x_n,t)= f_n (T_{-t+}^{(n)} (x_1, \ldots, x_n))
\end{equation}
for all $(x_1, \ldots, x_n) \in \Gamma_n^\ast$ and $f_n(t)=0$ on
$\Gamma_n\setminus \Gamma_n^\ast$, $n=1,2, \ldots$. If $f \in
\mathcal C$ then also $f(t) \in \mathcal C$ and the correlation
functions at time $t$ are still defined by (\ref{68}).
\begin{Theorem}\label{prop11}
Let $P$ be a signed measure on $\Gamma$ with density $f \in \mathcal
C.$ Then for every Borel set $\Delta\subset \Gamma_n$, $n=1,2,
\ldots$, the following identity holds
\begin{eqnarray}\label{71a}
&&\hspace{-20pt}\int_\Delta dx_1 \ldots dx_n \rho_n (x_1, \ldots,
x_n,t)\nonumber\\
&&\hspace{0pt} = \int_\Delta dx_1  \ldots dx_n \int_{\Delta (x_1,
\ldots, x_n;
[0,t])} d \delta W(x_1, \ldots, x_n, \delta)\nonumber\\
&&\hspace{20pt} \times\rho_{n+m(\delta)}(x_1(x_1, \ldots, x_n, \delta),
\ldots, x_{n+m(\delta)}(x_1, \ldots, x_n , \delta))\,.
\end{eqnarray}
The right (left) hand side of (\ref{71a}) does not depend on the
chosen version of $\rho$ $(\rho(t))$.
\end{Theorem}
\begin{Corollary}
Under the same assumptions, independently of the chosen versions of
$\rho(t)$ and $\rho$,
\begin{eqnarray}\label{72}
&&\hspace{-20pt}\rho_n (x_1, \ldots, x_n ,t)=
\int_{\Delta (x_1, \ldots, x_n, [0,t])} d \delta
W(x_1, \ldots, x_n, \delta)\nonumber\\&&\hspace{20pt}\times \rho_{n+m(\delta)} (x_1(x_1, \ldots,
x_n, \delta), \ldots, x_{n+m(\delta)} (x_1, \ldots, x_n ,
\delta))
\end{eqnarray}
$d x_1  \ldots  dx_n$ a.s..
\end{Corollary}
\begin{Proof}
Since the sum over $m$ is finite, we consider a term with fixed $m$
for reasons of notational simplicity. We abbreviate $x=(x_1, \ldots,
x_n)$. Then (\ref{71a}) reads
\begin{equation}\label{73}
\int_\Delta dx \rho_n(x,t) = \int_{\Delta^\ast} dx
\int_{\Delta(x;[0,t],m)} d\delta W(x,\delta) \rho_{n+m}
(y(x,\delta))
\end{equation}
with the obvious definition of $y(x,\delta) \in \Gamma_{n+m}^\ast$.
Here we replaced $\Delta$ by $\Delta^\ast = \Delta \cap
\Gamma_n^\ast$ which leaves the integral unchanged. Let $\mathcal
C^\ast \subset \mathcal C$ be the class of densities $f$ which
satisfy the continuity assumptions (iv) and (v), cf. (\ref{14b}),
omitting the requirements of normalization, positivity, and definite
number of particles. If $f\in \mathcal C^\ast$, then (\ref{73})
holds by Lemma \ref{prop9}. For an arbitrary $f \in \mathcal C$ there exists a
sequence $f^\varepsilon \in \mathcal C$ ´such that
\begin{equation}\label{74}
\lim_{\varepsilon \to 0} f^\varepsilon = f \quad \mbox{a.s.}
\end{equation}
on $\Gamma$.   Consequently
\begin{equation}\label{75}
\lim_{\varepsilon \to 0} \rho^\varepsilon = \rho \quad
\end{equation}
and, since
\begin{equation}\label{76}
\lim_{\varepsilon \to 0} f^\varepsilon (t) = f(t) \quad
\mbox{a.s.}\,,
\end{equation}
we also have
\begin{equation}\label{77}
\lim_{\varepsilon \to 0} \rho^\varepsilon (t) =\rho(t) \quad
\mbox{a.s.}\,.
\end{equation}
Therefore
\begin{equation}\lim_{\varepsilon \to 0} \int_\Delta dx \rho_n^\varepsilon(x,t) =
 \int_\Delta dx \rho_n(x,t)\,.
\end{equation}

We have to investigate now the convergence of the right hand side of
(\ref{73}). Let $\hat\rho \in {\mathcal C}^\ast$ and $\hat\rho = 0$
a.s.. Then also $\hat\rho (t) = 0$  a.s. and we conclude that
\begin{equation}\label{78}
0 = \int_{\Delta^\ast}d x \int_{\Delta(x;[0,t],m)} d \delta W(x,
\delta) \hat\rho_{n+m} (y(x, \delta))\,.
\end{equation}
Since all sets of measure zero can be approximated in this way, we
conclude that $y^{-1}$ (considered as a mapping for sets)  maps sets
in $\Gamma_{n+m}^\ast$ of $dx_1 \ldots dx_{n+m}$-measure zero to
sets in $\{(x, \delta) |x \in \Delta^\ast, \delta \in \Delta (x;
[0,t],m)\}$ of $dx d \delta$-measure zero.

By (\ref{75}) there exists a set $\widehat\Gamma
\subset\Gamma_{n+m}^\ast$ such that $\Gamma_{n+m} \setminus
\widehat\Gamma$ has measure zero and such that $\lim\limits_
{\varepsilon\to 0}
 \rho_{n+m}^\varepsilon = \rho_{n+m}$
pointwise on $\widehat\Gamma$. Let $\chi_{\widehat\Gamma}$ be the
indicator function of the set $\widehat\Gamma$. Then
\begin{equation}\label{79}
\lim_{\varepsilon \to 0} \rho_{n+m}^\varepsilon \chi_{\hat\Gamma}
(y(x, \delta)) = \rho_{n+m} \chi_{\hat\Gamma} (y(x,\delta))
\end{equation}
and by the argument given above

\begin{equation}\label{80}
\rho_{n+m}^\varepsilon (1- \chi_{\hat\Gamma}) (y(x,\delta)) = 0
\quad d x d \delta \quad \mathrm{a.s.}\,.
\end{equation}
Together with (\ref{64}) our claim follows from dominated
convergence.
\end{Proof}

%\vfill\eject

\end{document}